\documentclass[preprint,aps,showpacs,amsfonts,amssymb]{revtex4}
\usepackage{graphicx}

\begin{document}
\title{Surface Junction in Bi$_2$Sr$_2$CaCu$_2$O$_{8+d}$ Single Crystals}

\author{L. X. You}

\email[Email address: ]{lxyou@nju.edu.cn}

\affiliation{Research Institute of Superconductor Electronics,\\
             Department of Electronic Science and Engineering,\\
             Nanjing University, Nanjing 210093, China}

\author{P. H. Wu}

\affiliation{Research Institute of Superconductor Electronics,\\
             Department of Electronic Science and Engineering,\\
             Nanjing University, Nanjing 210093, China}

\begin{abstract}
As nature junction geometry of the $d$-wave cuprate superconductor/insulator/the mixture
of $s$-wave and $d$-wave cuprate superconductor, the upmost surface junction of intrinsic
Josephson junction stack fabricated on Bi$_2$Sr$_2$CaCu$_2$O$_{8+d}$ single crystals was
measured and discussed. We successfully observed half-integral microwave induced steps as
well as integral steps with resistively shunting technology, which shows the evidence of
second order supercurrent, consistent with Tanaka's theory.
\end{abstract}


\maketitle

Pairing symmetry in the cuprate superconductor is an important and controversial topic.
The recent development of phase-sensitive test, combined with the refinement of several
other symmetry-sensitive techniques, has for the most part settled this controversy in
favor of predominantly $d$-wave symmetry for a number of optimally hole- and
electron-doped cuprates \cite{Tsuei00}. Especially for Bi$_2$Sr$_2$CaCu$_2$O$_{8+d}$
(BSCCO), the phase-sensitive test results suggest that the $s$-wave component of the
order parameter should be vanishingly small \cite{Kirtley96}. Considering the crystal
structure of BSCCO, $s$-wave and $d$-wave pair states correspond to two distinct
irreducible representations and therefore are not allowed to mix \cite{Klemm98}. As a
result, BSCCO should be pure $d$-wave pairing symmetry. However, a few $c$-axis tunneling
experiments between BSCCO and $s$-wave superconductor Pb (BSCCO/Pb junction) consistently
unveiled the finite  supercurrent of first order. The results show $s$-wave component in
BSCCO should be small but nonzero \cite{Kleiner97,Mossle99}. However, the evidence of
$s$-wave is proven only in the upmost cuprate layer which couples to Pb. The question
raised from the contradiction is: Is the $s$-wave component intrinsic or induced from
$d$-wave state under some specific condition? Theoretical studies suggest that $s$-wave
paring symmetry may be induced in a d-wave superconductor, under certain conditions, by
spatial inhomogeneities such as surfaces, interfaces, or defects
\cite{Alvarez96,Zapotocky97,Martin98,Zhu98}. Reasonable explanation is that the $s$-wave
component of pairing state exists only in the surface cuprate layer of BSCCO, which is
induced from $d$-wave state by the surface effects.

According to a theory by Tanaka, second order $c$-axis pair tunneling is possible for a
junction between $c$-axis $s$-wave and pure $d$-wave superconductors \cite{Tanaka94}. In
other words, half-integer Shapiro steps should be observable in such a junction under
microwave irradiation. In the results of $c$-axis BSCCO/Pb junctions so far reported, the
observed supercurrent is dominantly of first order, and microwave induced Shapiro steps
only occur at voltages $V=nf\Phi_0$. Same results have also been observed in other
junctions of similar type \cite{Sun94,Kleiner96,Lesueur97,Woods99}. In essence, all the
junctions mentioned above are coupled by the $c$-axis pairing tunneling between $s$-wave
conventional superconductor and the cuprate superconductor with the mixture of $s$-wave
and $d$-wave pairing symmetry (s/I/s+$i$d), which was discussed theoretically in detail
\cite{Tanaka97}. Nevertheless, the $c$-axis pairing tunneling between the $d$-wave
cuprate superconductor and the cuprate superconductor with the mixture of $s$-wave and
$d$-wave pairing symmetry (d/I/s+$i$d) has not been discussed yet both in theory and
experiment. Natural geometry of d/I/s+$i$d junction, noteworthy but neglected, is the
upmost surface junction of intrinsic Josephson junction (IJJ) stack  fabricated on BSCCO
single crystals, which was discussed in a few papers without consideration of the mixture
of $s$-wave states and $d$-wave states \cite{Kim99,Doh00a,Doh00b}. In this letter, we
report and discuss the measurements of the surface junction and the observation of
microwave induced steps with resistively shunting method.

BSCCO single crystals were grown using floating zone technique in a four lamp arc-imaging
furnace. After cleaving from a bulk crystal, the fresh surface was deposited with a 70 nm
layer of gold to obtain a clean surface between single crystal and normal electrode as
well as to protect the surface from any contamination during further process of
fabrication and measurements. To make the contact resistance ohmic, the cleaved flake was
annealed in fresh O$_2$ flow for 10 minutes at 600$^0$C, then it was fixed on Si
substrate by Polyimide. With conventional photolithography and Ar ion etching, one mesa
with $a$-$b$ plane sizes of $6\times6\ \mu m^2$ covered with gold and photoresist was
formed. The junction number in the stack is controlled by the etching time and rate.
Finally a layer of SiO was deposited as insulator to protect and separate the mesa. After
lift-off, one electrode was connected to the top gold of the mesa, while two other
electrodes were connected to the ground plane of the mesa. All the transport measurements
described below were carried out with three-probe configuration.

\begin{figure}
\includegraphics{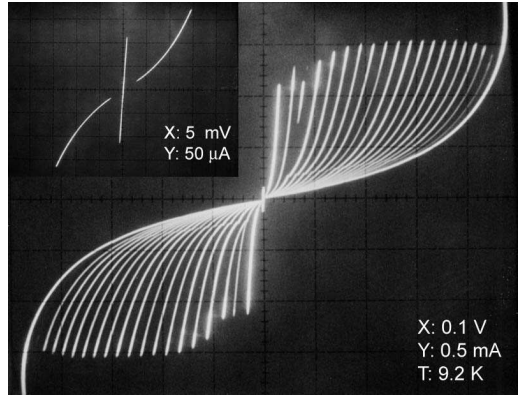} 
\caption{\label{fig:I-V} $I-V$ characteristics of the intrinsic Josephson stack
fabricated on BSCCO. The inset is the magnified figure of junction with the minimum
critical current I$_c'$.}
\end{figure}

In our experiments, the samples were measured in a liquid helium Dewar flask. Shown in
Fig.~\ref{fig:I-V} is the typical $I-V$ curve of the samples. The multiple quasi-particle
branches were developed with the increase of bias current. The junction number is counted
as 20 corresponding to the number of the branches. The critical current I$_c$ of most
junctions is nearly 1.5 mA at 9.2 K except one, the critical current I$_c'$ of which is
only 0.11 mA, one magnitude order smaller than the majority. The $I-V$ curve of this
novel junction is magnified and shown as the inset of main figure. Its supercurrent is a
little slantwise due to the linear contact resistance. This junction disappeared when the
temperature rose to T$_c'$=30K (35\%T$_c$, T$_c$ is the critical temperature of the bulk
crystal.) Doh etc. proved by progressively ion-beam etching experiment that the novel
junction is the upmost surface junction which forms by the upmost two cuprate layers
\cite{Doh00a}. The d-wave states in cuprate superconductor may induce to $s$-wave states
by spatial inhomogeneities such as surfaces, interfaces, or defects. As a result, the
pairing symmetry of the upmost cuprate layer changes from $d$-wave states to s+$i$d mixed
states with broken time-reversal symmetry due to the effect of surface. Moreover, the
superconductivity of the upmost cuprate layer is also suppressed. Consequently, the
critical temperature and critical current of the surface junction with the structure of
d/I/s+$i$d are greatly suppressed in comparison with the conventional d/I/d IJJ.

\begin{figure}
\includegraphics{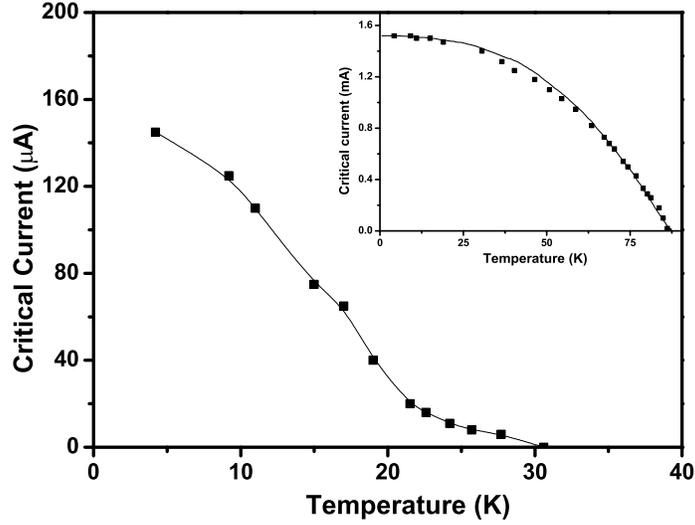} 
\caption{\label{fig:Ic-T} $I_c-T$ characteristics of the surface junction. The inset is
$I_c-T$ of conventional intrinsic junction, the solid line in which is theoretical
calculation results for $c-$axis d/I/d junction.}
\end{figure}

The characteristics of critical current of the surface junction varied with temperature
$I_c-T$ was measured and shown in Fig.~\ref{fig:Ic-T}. As comparison, $I_c-T$ of
conventional IJJ is also shown as the inset of Fig.~\ref{fig:Ic-T}. the solid line in the
inset is theoretical calculation of $I_c-T$ for $c$-axis d/I/d junction by Tanaka etc
\cite{Tanaka97}. The experimental result is consistent with it. However, the evident
deviation was found in the surface d/I/s+$i$d junction. It can be fitted by abnormal
temperature dependence $I_c\varpropto(T_c-T)^2$ instead of $I_c\varpropto T_c-T$ when the
temperature is close to Tc.

\begin{figure}
\includegraphics{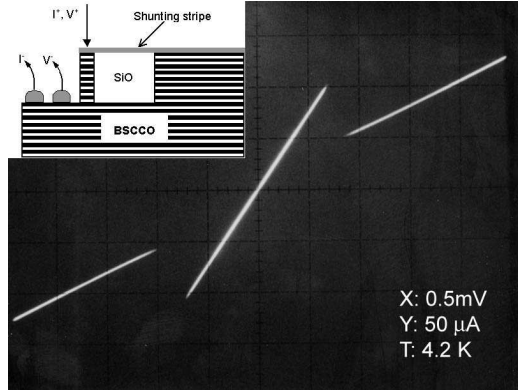} 
\caption{\label{fig:RSIJJ} I-V characteristics of resistively shunted surface junction.
the inset shows the schematic figure of resistively shunted intrinsic Josephson
junction.}
\end{figure}

\begin{figure}
\includegraphics{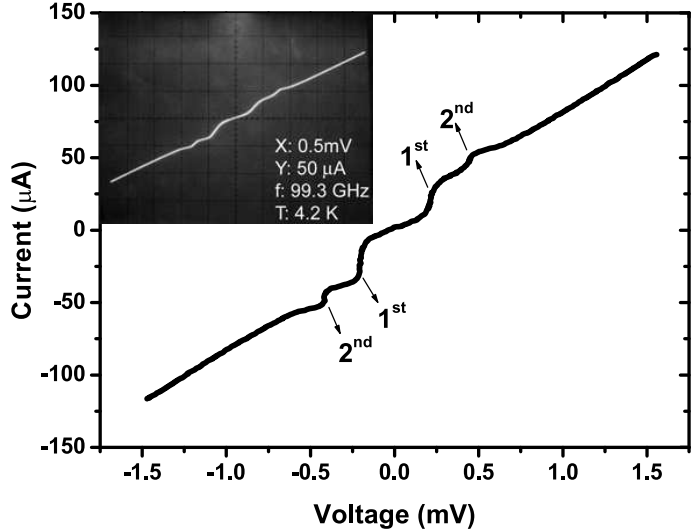} 
\caption{\label{fig:Step1} Integer microwave induced steps of resistively shunted surface
junction with the microwave irradiation at 99.3 GHz, the contact resistance is
abstracted. the inset shows the original image.}
\end{figure}

\begin{figure}
\includegraphics{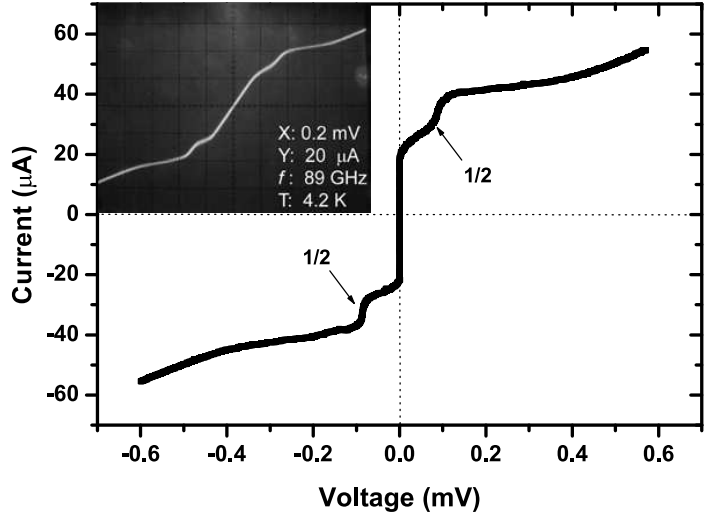} 
\caption{\label{fig:Step2} half-integer microwave induced steps of resistively shunted
surface junction with the microwave irradiation at 89 GHz, the contact resistance is
abstracted. The inset shows the original image.}
\end{figure}

To investigate the detailed feature of the surface d/I/s+$i$d junction, the microwave
response was measured. However, with microwave irradiation at $80\sim100$ GHz and lower
frequency, no microwave-induced steps were observed. Because Shapiro steps are only
observable when $f_{rf}\gg f_p$ in heavy under-damped Josephson junction
\cite{Kautz81,Wang01a}. And the plasma oscillation frequency $f_p$ of the surface
junction is estimated to be 100 GHz from $f_p=(\frac{eJ_cd}{\pi
h\varepsilon_r\varepsilon_0})^{1/2}$, where the critical current density $J_c$ is about
300 A/cm$^2$, the thickness of inter-layer d is 12 \AA~for BSCCO, and the relative
dielectric constant $\varepsilon_r$ is 3.5 \cite{Walkenhorst97}. To observe
microwave-induced steps, resistively shunted technology was adopted. It has been well
established that Shapiro steps can be observable in resistively shunted IJJ
\cite{Wang00,Wang01b}. In this letter, we use a gold stripe to shunt the mesa, the inset
of Fig.~\ref{fig:RSIJJ} is the schematic figure of the resistively shunted IJJ. For the
critical current of the surface junction is lowest in the stack, the surface junction
will be firstly shunted. Detailed discussion of resistively shunting theory was published
elsewhere \cite{Wang01b}. I-V curves shown in Fig.~\ref{fig:RSIJJ} shows typical behavior
of RSJ model with a linear contact resistor about $6~\Omega$ in series. The resistance is
subtracted in the following figures. With microwave irradiation at 99.3 GHz, the
Microwave-induced steps up to second order were observed and shown in
Fig.~\ref{fig:Step1}. The interval between the first order steps is 410 $\mu$V, which
fits the Josephson voltage-frequency relationship $V=f\Phi_0$ very well. It proves the
microwave-induced steps are of the first order supercurrent, which derives from the pair
coupling between $d$-wave component in the surface layer and inner $d$-wave layer.
Fig.~\ref{fig:Step2} shows the microwave response of the resistively shunted surface
junction when the frequency of microwave irradiation changed to 89 GHz. Only two steps
were obtained, the interval between them is about 185 $\mu$V, which fits exactly at
$V=0.5f\Phi_0$. The half-integral steps prove the existence of the second order
supercurrent, which originates from the pair coupling between $s$-wave component in the
surface layer and $d$-wave in the inner cuprate layer. It is consistent with the
prediction of Tanaka \cite{Tanaka94}. When the power of microwave source increases, the
half-integral steps disappear and conventional Shapiro steps appear. It suggests that the
second order supercurrent is more sensitive to microwave irradiation than first order
supercurrent derived from $c$-axis pair tunneling between d-wave states. However, because
of the low content of $s$-wave states in the mixture s+$i$d states, the microwave induced
half-integral steps were concealed by the conventional Shapiro steps with the increase of
the microwave power.

In summary, as natural geometry of d/I/s+$i$d junction, the surface junction in intrinsic
Josephson junctions stack was measured and discussed. The temperature dependence of
critical current shows a significantly difference from the intrinsic d/I/d Josephson
junction and follows abnormal temperature dependence $I_c\varpropto (Tc-T)^2$ when the
temperature is close to T$_c$. With resistively shunting method, the surface junction
shows typically RSJ model, the half-integral microwave induced steps were observed as
well as conventional integral Shapiro steps, which shows the existence of second order
supercurrent and supports strongly Tanaka's theory.

We thank W. Y. Zhang, J. Li for valuable discussions, H. B. Wang for his technical
instruction. This work was supported by the Ministry of Science and Technology of China
(Grant No. G19990646), partially carried out at Laboratory of Electronic Intelligent
Systems, Research Institute of Electronic Communication, Tohoku University, Japan.


\begin{thebibliography}{50}
\bibitem{Tsuei00}C. C. Tsuei and J. R. Kirtley, Rev. Mod. Phys. \textbf{72}, 969
(2000)
\bibitem{Kirtley96}J. R. Kirtley \textit{et al.}, Europhys. Lett.
\textbf{36}, 707 (1996)
\bibitem{Klemm98}R. A. Klemm, C. T. Rieck, and K. Scharnberg, Phys. Rev. B \textbf{58}, 1051 (1998)
\bibitem{Kleiner97}R. Kleiner \textit{et al.}, Physica C
\textbf{282}, 2435 (1997)
\bibitem{Mossle99}M. M\"ossle and R. Kleiner, Phys. Rev. B \textbf{59}, 4486 (1999)
\bibitem{Alvarez96}J. J. Vicente Alvarez, G. C. Buscaglia, and C. A. Balseiro, Phys. Rev. B \textbf{54}, 16168 (1996)
\bibitem{Zapotocky97}M. Zapotocky, D. L. Maslov, and P. M. Goldbart, Phys. Rev. B \textbf{55}, 6599 (1997)
\bibitem{Martin98}A. M. Martin and J. F. Annett, Phys. Rev. B \textbf{57}, 8709 (1998)
\bibitem{Zhu98}J. X. Zhu and C. S. Ting, Phys. Rev. B \textbf{57}, 3038 (1998)
\bibitem{Tanaka94}Y. Tanaka, Phys. Rev. Lett. \textbf{72}, 3871 (1994)
\bibitem{Sun94}A. G. Sun, D. A. Gajewski, M. B. Maple, and R. C. Dynes, Phys. Rev. Lett. \textbf{72}, 2267 (1994)
\bibitem{Kleiner96}R. Kleiner \textit{et al.}, Phys. Rev. Lett. \textbf{76},(1996) 2161
\bibitem{Lesueur97}J. Lesueur \textit{et al.}, Phys. Rev. B \textbf{55} (1997) 3398
\bibitem{Woods99}S. I. Woods \textit{et al.}, IEEE Trans, Appl. Supercond. \textbf{9}, (1999) 3917
\bibitem{Tanaka97}Y. Tanaka and S. Kashiwaya, Phy. Rev. B \textbf{56}, 892 (1997)
\bibitem{Kim99}N. Kim, Y. -J. Doh, H. -S. Chang, and H. -J. Lee, Phys. Rev. B \textbf{59}, 14639 (1999)
\bibitem{Doh00a}Y. -J. Doh, H. -J. Lee, and H. -S. Chang, Phys. Rev. B \textbf{61} 3620 (2000)
\bibitem{Doh00b}Y. -J. Doh, J. Kim, K. -T. Kim, and H. -J. Lee, Phys. Rev. B \textbf{61}, R3834 (2000)
\bibitem{Kautz81}R. L. Kautz, J. Appl. Phys. \textbf{52}, 3528 (1981)
\bibitem{Wang01a}H. B. Wang, P. H. Wu, and T. Yamashita, Phys. Rev. Lett. \textbf{87}, 107002 (2001)
\bibitem{Walkenhorst97}W. Walkenhorst \textit{et al.}, Phys. Rev. B \textbf{56}, 8396 (1997)
\bibitem{Wang00}H. B. Wang \textit{et al.}, Appl. Phys. Lett. \textbf{77}, 1017 (2000)
\bibitem{Wang01b}H. B. Wang \textit{et al.}, Physica C \textbf{362}, 108 (2001)
\end{thebibliography}
\end{document}